\documentclass[9pt,twocolumn,twoside]{pnas-new}

\usepackage[version=3]{mhchem}

\templatetype{pnasresearcharticle} 
\setboolean{displaywatermark}{false}

\begin{document}

\title{Fundamental Limits of Dissociative Electrochemical Ammonia Synthesis via Electrodeposited Metals}

\author[a]{Victor Azumah}
\author[a,b,c,1]{Venkatasubramanian Viswanathan}

\affil[a]{Department of Chemical Engineering, University of Michigan, Ann Arbor, MI 48109}
\affil[b]{Department of Aerospace Engineering, University of Michigan, Ann Arbor, MI 48109}
\affil[c]{Department of Mechanical Engineering, University of Michigan, Ann Arbor, MI 48109}

\leadauthor{Azumah}

\authorcontributions{
V.A. and V.V. conceived the study. V.A. derived the thermodynamic framework and performed all calculations with input from V.V. All authors contributed to data analysis, interpretation, and writing of the manuscript.
}

\authordeclaration{
V.V. is an inventor on a provisional patent application related to proton donors for lithium-mediated ammonia synthesis.
}

\correspondingauthor{\textsuperscript{1} Venkatasubramanian Viswanathan. E-mail: venkvis@umich.edu}

\begin{abstract}
Electrochemical ammonia synthesis via lithium-mediated nitrogen dissociation has demonstrated exceptional Faradaic efficiency at ambient conditions, but its viability is limited by a high energy cost of $\sim$9.12 eV per \ce{NH3} via lithium electrodeposition. Here, we establish the thermodynamic limits for dissociative nitrogen reduction using elemental metals by decomposing the process into three steps: metal deposition, nitridation, and protonation. We derive energetic constraints that any viable mediator must satisfy and show that highly reducing metals impose significant energetic penalties. To reduce this cost, we explore solvent tuning and bimetallic alloy strategies that shift deposition potentials without compromising nitridation spontaneity. Our results offer design principles for lowering the energy input of dissociative nitrogen reduction while maintaining its selectivity advantage over associative routes.

\end{abstract}

\dates{This manuscript was compiled on \today}
\doi{\url{www.pnas.org/cgi/doi/10.1073/pnas.XXXXXXXXXX}}

\maketitle
\thispagestyle{firststyle}
\ifthenelse{\boolean{shortarticle}}{\ifthenelse{\boolean{singlecolumn}}{\abscontentformatted}{\abscontent}}{}

\dropcap{A}mmonia has grown to become one of the top commodity chemicals in the world due to its role as crucial component in agricultural fertilizer. Produced by the reaction of hydrogen and nitrogen gas in the energy intensive Haber-Bosch process, it currently consumes 1-2\% of the world's energy, using it to provide the harsh temperatures and pressures to break the $\ce{N_{2}}$ triple bond. Given these conditions and the sourcing of hydrogen gas from catalytic cracking, the process also consumes 2-3\% of the global natural gas supply, producing with it 1.9 - 9.3 ton of \ce{CO_{2}} per ton of \ce{NH_{3}}. This makes \ce{NH_{3}} production one of the highest sources of greenhouse gas emissions\cite{lazouski2019understanding}.

The rediscovery of the lithium-mediated process pioneered by Tsuneto et al., which takes advantage of lithium's ability to break the strong nitrogen bond at ambient conditions, has allowed for electrochemical reduction of nitrogen through a dissociative pathway\cite{tsuneto1994lithium}. Since 2019, there have been rapid improvements in Faradaic efficiencies of the lithium-mediated process, increasing from 8.4\% to consistently above 60\%, with some work even being able to achieve 100\% single-cycle efficiencies\cite{suryanto2021nitrogen,du2022electroreduction}. These improvements have come from increasing understanding of solid-electrolyte interphase (SEI) formation, morphology and their effects on reactant transport; proton donor effects on SEI morphology\cite{steinberg2023imaging}, and optimizations in solvent and salt components\cite{li2022electrosynthesis,iriawan2024upshifting}. Recent work has expanded the successes of the lithium-mediated process to other ionic nitride metals such as magnesium and calcium, also achieving efficiencies above 40\%\cite{fu2024calcium,goyal2024metal}. The success of these main group metals primarily stem from their ability to split the nitrogen triple bond in a nitride formation process, allowing them to take advantage of a kinetically facile dissociative pathway\cite{lazouski2019understanding,goyal2024metal,fu2024calcium}. However, this comes at the cost of electrodepositing such metals at their highly negative reduction from solution to provide the surface for spontaneous nitride formation\cite{zhou2024electrochemical}, with lithium, the poster child of the dissociative process, requiring a 9.12eV (879.kJmol$^{-1}$) minimum energy input for its nitrogen reduction process.

In Zhou et al's recent analysis, they suggested the circumvention of the high reductive costs on metals like Li and Ca by returning to transition metals like Fe or Ru, or increasing the temperature of the electrochemical cell\cite{zhou2024electrochemical}. The question remains whether the dissociative scheme for nitrogen reduction is reasonable given the energy cost, or if a return to the associative scheme of \ce{N2} reduction, where a more dominant hydrogen evolution reaction keeps selectivities low\cite{singh2019strategies}, is in order. In this report, we assess the viability of elemental metals within the dissociative scheme as replacements for Li and suggest ways for reducing the Li's energetic cost within the dissociative scheme.

\section{Results and discussion}
\subsection{The Free Energy criterion for Mediator Choice}
\begin{figure*}[hbt!]
\begin{center}
\includegraphics[width=17.8cm]{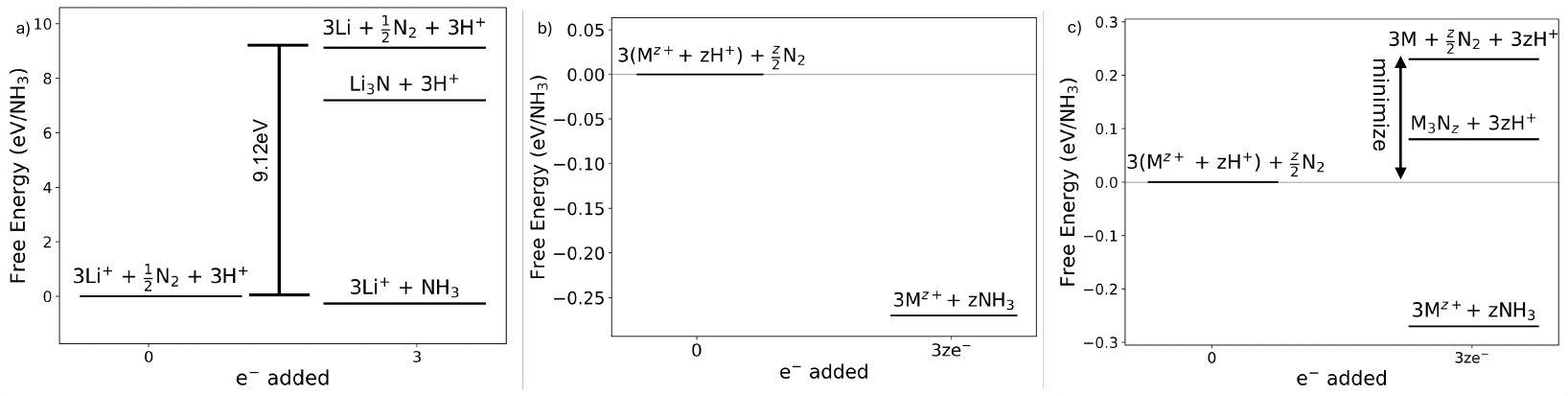}
\caption{Free energy diagrams showing the different steps of ammonia reduction via (a) the high energy cost associated with electrodepositing lithium in the current nitride-mediated scheme, (b) an ideal electrocatalytic mediator requiring no energy input, (c) a real nitride-mediated catalyst which minimizes energy input. A replacement for Li as an electrocatalyst for this process would need to minimize the energy input whilst ensuring that the nitride formation step and subsequent protolysis so \ce{NH3} remain downhill.}
\label{fig:MxNySteps}
\end{center}
\end{figure*}

We start our analysis by defining electrochemical dissociative nitrogen reduction. The generally agreed mechanism for this dissociative pathway shown in Fig.~\ref{fig:MxNySteps}a, starts with (i) electrodeposition of the metal ions on an electrode (Eq.~\ref{eqnMdep}), (ii) nitrogen dissociation through the formation of a metal nitride (Eq.~\ref{eqnMxNy}), and finally (iii) formation of ammonia by protonation of the nitride(Eq.~\ref{eqnMtoNH3}).

Restricting our analysis to the case of trivalent nitrides where the nitrogen oxidation number is the same as in ammonia (-3), the equivalent steps are:
\begin{equation}\label{eqnMdep}
    3(\mathrm{M}^{z+} + \mathrm{ze}^{-}) \rightarrow 3\mathrm{M(s)}
\end{equation}
\begin{equation}\label{eqnMxNy}
    \mathrm{3M}(s) \, + \mathrm{\frac{z}{2}N_{2}}(g) \rightarrow \mathrm{M_{3}N_{z}}(s)
\end{equation}
\begin{equation}\label{eqnMtoNH3}
    \mathrm{M_{3}N_{z}}(s) + \mathrm{3zH}^{+} \rightarrow \mathrm{3M}^{z+} + \mathrm{zNH_{3}(g)}
\end{equation}

The net reaction is the electrochemical reduction of the nitrogen molecule via protons.
\begin{equation}\label{eqnN2NH3}
    \mathrm{\frac{z}{2}N_{2}}(g) + \mathrm{3z(H^{+} + e^{-})}\rightarrow \mathrm{zNH_{3}(g)}
\end{equation}
From  Eq.~\ref{eqnMtoNH3}, we observe that for this specific case where a trivalent nitride is produced, there is no electron transfer accompanying the protonation reaction. The only electrochemical step is the electrodeposition step in which electrons are used to raise the electrochemical potential of the metal, allowing it to reduce nitrogen in the subsequent nitride formation step, and providing the energy for the protonation step, both of which are chemical. This can be seen from the free energy diagram of the lithium-mediated process (Fig.~\ref{fig:MxNySteps}a).

We start by considering the free energy of \ce{NH_{3}} production per \ce{NH_{3}} via metal nitride formation through its three steps. For the metal deposition step Eq.~\ref{eqnMdep}, this is simply calculated from its deposition potential.
\begin{equation}\label{Gdep}
    \mathrm{\Delta G_{1}}= -3FE_{M^{z+}||M}
\end{equation}
where z is the charge on the metal ion. At the nitride step Eq.~\ref{eqnMxNy}, the free energy is simply the Gibbs energy of formation of the metallic nitride as its formed from the component elements M and \ce{N_{2}}.
\begin{equation}\label{GMxNy}
    \mathrm{\Delta G_{2}}= \frac{\Delta G_{f}(M_{3}N_{z})}{z}
\end{equation}
and finally the formation of \ce{NH_{3}} from the produced nitride film Eq.~\ref{eqnMtoNH3}. Here the free energy of the reaction has the general expression:
\begin{equation}\label{GNH3vers1}
    \mathrm{\Delta G_{3}}= \Delta G_{f}(NH_{3}) + \frac{3\Delta G_{M^{z+}}}{z} -3\Delta G_{H^{+}} -\frac{\Delta G_{f}(M_{3}N_{z})}{z}
\end{equation}
Using the standard hydrogen electrode, we deduce that the formation energy of metallic ions less that if the protons is simply the reverse reduction potential of the metal referenced to the standard hydrogen electrode. We rewrite Eq.~\ref{GNH3vers1} as:
\begin{equation}\label{GNH3vers2}
    \mathrm{\Delta G_{3}}= \Delta G_{f}(NH_{3}) + 3FE_{M^{z+}||M} -\frac{\Delta G_{f}(M_{3}N_{z})}{z}
\end{equation}
and again rewrite that in terms of the previous free energies:
\begin{equation}\label{GNH3vers3}
    \mathrm{\Delta G_{3}}= \Delta G_{f}(NH_{3}) -\mathrm{\Delta G_{1}} -\mathrm{\Delta G_{2}}
\end{equation}
By Hess's law, we require that the sum of free energies of all three steps Eqs.(\ref{eqnMdep}-\ref{eqnMtoNH3}) be equal to the overall reaction step Eq.~\ref{eqnN2NH3} which means that the net free energy (sums of Eqs.~\ref{Gdep}-\ref{GNH3vers1}) must be the free energy for \ce{NH_{3}} formation ($\Delta G_{f}(NH_{3})$).

\begin{figure*}[tb]
\begin{center}
\includegraphics[width=17.8cm]{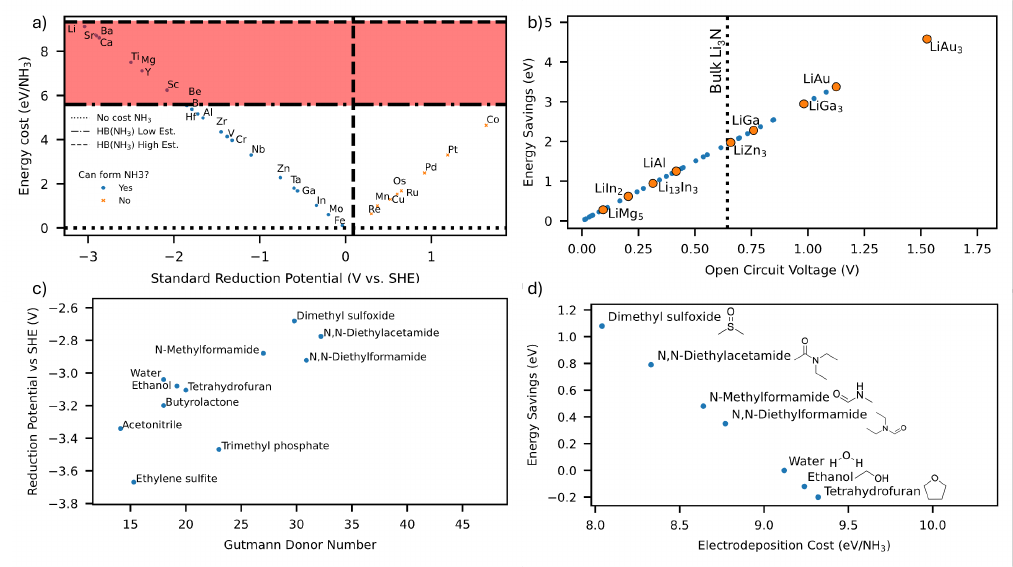}
\caption{Thermodynamic Energy cost of producing \ce{NH_{3}} using different mediators assuming electrodeposition and chemical nitridation.  Comparison to high\cite{singh2019strategies} and low\cite{macfarlane2020roadmap} estimates for Haber-Bosch \ce{NH_{3}} show that even at their thermodynamic limit, favored catalysts for nitride-mediated ammonia synthesis (Li, Ca) can barely achieve parity with more inefficient HB systems. Lower costs elements like Fe and Mo have been limited by their selectivity in electrochemical nitrogen reduction\cite{singh2019strategies}. Viable design towards lower cost dissociative NRR requires (b) Search for bimetallic alloys of Li to reduce electrodeposition cost and (c) Finding solvents that reduce the Li electrodeposition potential with (d) cost savings of Li electrodeposition in some favorable solvents. The exploration of these strategies and their combinations could significantly reduce the cost of dissociative NRR with Li.}
\label{fig:Solutions}
\end{center}
\end{figure*}
Only the first step of the reaction - metal deposition, is an electrochemical step, the electrons raising the chemical potential of the metal to provide energy for the subsequent nitride formation and protonation steps, both of which are purely chemical. We deduce from Eqs.~\ref{eqnMxNy} and ~\ref{eqnMtoNH3} that the energy for the chemical steps come from the conversion of the metal into its ions, establishing our first condition. The dissolution potential from the metal to its ions should thus be positive relative to the potential for reducing nitrogen to \ce{NH_{3}}. Stated alternatively, the reduction potential of the metal should be more negative than the nitrogen reduction potential. The cell potential for nitrogen reduction has been found to be 0.09V vs SHE\cite{viswanathan2014unifying}. From this condition, we are able to eliminate any metals with reduction potentials more positive than 0.09V including more noble metals like Au and Pt, whose conversion from metal to ions would make them far too stable, hindering their ability to spontaneously fund the chemical steps that follow.

With our first condition established and the knowledge that the second two steps are purely chemical, we require that both steps be in themselves spontaneous so that they can occur without any additional energy input as this would result in increased inefficiencies within the system. This provides our second and third conditions - that $\Delta G_{2}$ and $\Delta G_{3}$ must be less than or equal to zero. We can combine our three criteria for the free energies with Hess's law of heat summation to obtain:
\begin{equation}\label{G1_constrained}
    \mathrm{\Delta G_{1}} - |\mathrm{\Delta G_{2}}| - |\mathrm{\Delta G_{3}}| \geq \Delta G_{f}(NH_{3}) 
\end{equation}
We see that Eq.~\ref{G1_constrained} is similar to Eq.~\ref{GNH3vers3} subject to the constrains defined for each step. Applying the Eqs.~\ref{GMxNy} and \ref{GNH3vers3} to our constrained system, we are able to obtain our final criterion:
\begin{equation}\label{FinalConstraint}
    \mathrm{\Delta G_{1}} \geq  \Delta G_{f}(NH_{3}) 
\end{equation}

We calculate the energy cost for producing \ce{NH_{3}} with a mediator simply by taking the sum of our energy input steps. For mediators observing the imposed constraints, this is just the energy of the electrodeposition step, given by:
\begin{equation}\label{Efficiency}
     \mathrm{Energy \, Cost \,(eV/NH_{3})} = \mathrm{E_{input}} = \Delta G_{1}
\end{equation}
where $E_{input}$ is the sum of all endothermic steps. The second equality is only valid for mediators satisfying the constraints we have detailed. In the perfect mediator, there is a single electrochemical step from 0V vs SHE to the \ce{NH_{3}} formation potential (0.09V), allowing us to do work whilst forming \ce{NH3}\cite{viswanathan2014unifying} (Fig.~\ref{fig:MxNySteps}b). In a real process however, the best mediator is one whose deposition potential is slightly more negative than but as close to that \ce{NH_{3}} formation potential as possible, minimizing the energy required to plate the metal and make the chemical steps downhill ($\Delta G_{1}$) (see Fig.~\ref{fig:MxNySteps}c).

\subsection{Mediator Choice Effects on Energy Cost}
With the criteria established, we explore a series of known metallic species, using how their reduction potential, nitride formation energies, and protonation energies ($\Delta G_{3}$) to determine whether or not they are able to form \ce{NH_{3}} and calculating their energy costs from Eq.~\ref{Efficiency}. 

In Fig.~\ref{fig:Solutions}a, we consider the relationship between standard reduction potential and energy cost per \ce{NH_{3}} according to the reaction mechanism detailed in Eqs.~\ref{eqnMdep}-~\ref{eqnMtoNH3}. For mediators that are more negative than the \ce{N_{2}} reduction potential  (0.09V), we observe a linear relationship consistent with Eq.~\ref{Efficiency}. For those mediators, the lowest energy costs are associated with the weakly reducing metals such as Fe and Mo. We find that past a reducing potential of -2V, the energy cost for the metal-mediated process becomes less favorable than the best Haber-Bosch estimates\cite{macfarlane2020roadmap}, with highly reducing metals such as Li, Ba, Sr, having the most cost intensive \ce{NH_{3}} production costs. For mediators more positive than the \ce{N_{2}} potential, the energy costs tend to be much lower, though the spontaneous deposition step makes the chemical steps endothermic and uphill, preventing them from producing \ce{NH_{3}} in this scheme.

In a system where electrochemical nitridation occurred i.e. Eqs.~\ref{eqnMdep}-~\ref{eqnMxNy} happen simultaneously, the energy input for conversion to the metal is however offset by the energy released in the nitridation step. Thus, we see that the total energy cost for some mediators more negative than \ce{N_{2}} reduction potential decreases, with the magnitude of this reduction given by Eq.~\ref{eqnMxNy}. We find that even with this offset, highly reducing metals such as Li and Ca remain very energy intensive, still having higher \ce{NH_{3}} costs than the best HB estimates.

Though we cannot assess the viability or discuss design towards an electrochemical nitridation step, we highlight that the same functional step can be achieved by reducing the Li electrodeposition potential to that of the nitride formation step. One way to do this is through the use of bimetallic lithium alloys (Fig.~\ref{fig:Solutions}b). Energy savings compared to lithium would require that the alloy Li-M forms below Li electrodeposition potentials. Assuming a nitride formation step through \ce{Li3N} as opposed to \ce{M_xN_z} where M is the alloying metal, we would additionally require that the open-circuit voltage for forming the Li-M alloy is not too positive, lest the nitride formation step become endothermic and uphill. This design requires simply that $-3\times OCV(Li-M)\leq -1.93$, the formation energy of \ce{Li3N} or  $OCV(Li-M)\leq 0.6433$. Whilst Li still remains thermodynamically expensive even at the nitride formation energy, there are gains to be had for other metals like Mg and Al who have more room to maneuver within this design space due to their higher nitride formation energies, -4.71eV and -3.36eV, respectively. Of course, the formation ternary nitrides \ce{Li_{a}M_{b}N_{z}} makes it possible to widen or close this design window, but we reserve that analysis for future work.

\subsection{Effects on Electrolyte Choice on Energy Cost}
Drawing inspiration from the field of Li-air batteries\cite{khetan2015trade} and recent work within the field\cite{iriawan2024upshifting}, we point to the solution of shifting the Li electrodeposition potential by tuning the electrolyte from which Li is electrodeposited. Using the converted half-wave potentials from existing work\cite{gritzner2010standard}, it is clear that several solvents exists within which the Li electrodeposition potential is much lower than in (-3.04V vs SHE) and than in tetrahydrofuran (-3.11V vs SHE), the current standard solvent used in the lithium-mediated process (Fig.~\ref{fig:Solutions}c). Of those identified from previous work, we note that up to 1.3eV can be saved from the electrodeposition cost alone assuming negligble effects of the solvent on the \ce{NH3} solvation. As the formation of nitride occurs with respect to bulk Li, the gains from electrodeposition this way is not constrained to the nitride formation energy and not exclusive to the design of bimetallic alloys outlined above, with gains of up to 3eV from a possible synergistic process combining the two (Fig.~\ref{fig:Solutions}d). Again, this highlights that pathways to lower cost dissociative nitrogen reduction potential may exist for reducing metals with lower electrodeposition costs and activation energy barriers comparable to Li, like Ca and Cr\cite{zhou2024electrochemical}.

\subsection{Conclusions}
In this work, we assessed the problem of high cost dissociative nitrogen reduction by exploring the thermodynamics feasibility of elemental metals as replacements for lithium. We show that appropriate mediators for electrochemical \ce{NH_{3}} synthesis via a nitride (\ce{M_{3}N_{z}}) intermediate require an electrodeposition step more negative than the \ce{N_{2}} reduction potential (0.09V) and that this step must provide enough energy to make the subsequent chemical steps spontaneous. We find that the energy cost to produce \ce{NH_{3}} associated with highly reducing metals like Li, Ca, and Mg, owing to their highly electrodeposition potentials, makes them less favorable to current Haber-Bosch \ce{NH_{3}}, even before accounting for process overpotentials. We show that through a combination of alloy and electrolyte design, reductions in the cost of a dissociative process can be achieved solely by reducing the electrodeposition cost. There remains the engineering problem of finding the right combination of electrolytes and surface. But our analysis in the very least shows that there remains a path to lower cost dissociative nitrogen reduction.

\acknow{The authors would like to acknowledge Lance Kavalsky and Archie Mingze Yao for being sounding boards for important discussions regarding insights and calculations.}

\showacknow{} 


\bibliography{ref}

\end{document}